# Comparison of machine learning algorithms for merging gridded satellite and earth-observed precipitation data

Georgia Papacharalampous[1,*], Hristos Tyralis[2], Anastasios Doulamis[3], Nikolaos Doulamis[4]

[1] Department of Topography, School of Rural, Surveying and Geoinformatics Engineering, National Technical University of Athens, Iroon Polytechniou 5, 157 80 Zografou, Greece (papacharalampous.georgia@gmail.com, https://orcid.org/0000-0001-5446-954X)

[2] Department of Topography, School of Rural, Surveying and Geoinformatics Engineering, National Technical University of Athens, Iroon Polytechniou 5, 157 80 Zografou, Greece (montchrister@gmail.com, hristos@itia.ntua.gr, https://orcid.org/0000-0002-8932-4997)

[3] Department of Topography, School of Rural, Surveying and Geoinformatics Engineering, National Technical University of Athens, Iroon Polytechniou 5, 157 80 Zografou, Greece (adoulam@cs.ntua.gr, https://orcid.org/0000-0002-0612-5889)

[4] Department of Topography, School of Rural, Surveying and Geoinformatics Engineering, National Technical University of Athens, Iroon Polytechniou 5, 157 80 Zografou, Greece (ndoulam@cs.ntua.gr, https://orcid.org/0000-0002-4064-8990)

* Corresponding author

**Abstract**: Gridded satellite precipitation datasets are useful in hydrological applications as they cover large regions with high density. However, they are not accurate in the sense that they do not agree with ground-based measurements. An established means for improving their accuracy is to correct them by adopting machine learning algorithms. This correction takes the form of a regression problem, in which the ground-based measurements have the role of the dependent variable and the satellite data are the predictor variables, together with topography factors (e.g., elevation). Most studies of this kind involve a limited number of machine learning algorithms, and are conducted for a small region and for a limited time period. Thus, the results obtained through them are of local importance and do not provide more general guidance and best practices. To provide results that are generalizable and to contribute to the delivery of best practices, we here compare eight state-of-the-art machine learning algorithms in correcting satellite precipitation data for the entire contiguous United States and for a 15-year period. We use monthly data from the PERSIANN (Precipitation Estimation from Remotely Sensed Information using Artificial Neural Networks) gridded dataset, together with monthly earth-observed precipitation data from the Global Historical Climatology Network monthly database, version 2 (GHCNm). The results suggest that extreme gradient boosting (XGBoost) and random forests are the most accurate in terms of the squared error scoring function. The remaining algorithms can be ordered as follows from the best to the worst: Bayesian regularized feed-forward neural networks, multivariate adaptive polynomial splines (poly-MARS), gradient boosting machines (gbm), multivariate adaptive regression splines (MARS), feed-forward neural networks, linear regression.

**Keywords**: benchmarking; big data; gradient boosting machines; PERSIANN; poly-MARS; random forests; remote sensing; satellite data correction; spatial interpolation; XGBoost

## 1. Introduction

Knowing the quantity of precipitation at a dense spatial grid and for an extensive time period is important in solving a variety of hydrological engineering and science problems, including many of the major unsolved problems listed in Blöschl et al. (2019). The main sources of precipitation data are ground-based gauge networks and satellites (Sun et al. 2018). Data from ground-based gauge networks are precise; however, maintaining such a network with high spatial density and for a long time period is costly. On the other hand, satellite precipitation data are cheap to obtain but not accurate (Mega et al. 2019, Salmani-Dehaghi and Samani 2021, Li et al. 2022, Tang et al. 2022).

By merging gridded satellite precipitation products and ground-based measurements, we can obtain data that are more accurate than the raw satellite data and, simultaneously, cover space with a much higher density compared to the ground-based measurements. This merging is practically a regression problem in a spatial setting, with the satellite data being the predictor variables and the ground-based data being the dependent variables.

Such kinds of problems are also commonly referred to under the term "downscaling" and are special types of spatial interpolation. The latter problem is met in a variety of fields (see, e.g., the reviews by Bivand et al. 2013, Li and Heap 2014, Heuvelink and Webster 2022, Kopczewska 2022). Reviews of the relevant methods for the case of precipitation can be found in Hu et al. (2019) and Abdollahipour et al. (2022).

Spatial interpolation of precipitation by merging satellite precipitation products and ground-based measurements has been conducted at multiple temporal and spatial time scales by using a variety of regression algorithms, including several machine learning ones. A non-exhaustive list of previous works on the topic and a summary of their methodological information can be found in Table 1. Notably, this table is indicative of the large diversity in the temporal and spatial scales examined and in the algorithms utilized.

Table 1. Summary of previous studies and the present study on merging gridded satellite precipitation products and ground-based measurements.

| Study | Time scale | Spatial scale | Algorithms |
|---|---|---|---|
| He et al. (2016) | Hourly | South-western, central, north-eastern and south-eastern United States | Random forests |
| Meyer et al. (2016) | Daily | Germany | Random forests, artificial neural networks, support vector regression |
| Tao et al. (2016) | Daily | Central United States | Deep learning |
| Yang et al. (2016) | Daily | Chile | Quantile mapping |
| Baez-Villanueva et al. (2020) | Daily | Chile | Random forests |
| Chen et al. (2020a) | Daily | Dallas–Fort Worth in the United States | Deep learning |
| Chen et al. (2020b) | Daily | Xijiang basin in China | Geographically weighted ridge regression |
| Rata et al. (2020) | Annual | Chéliff watershed in Algeria | Kriging |
| Chen et al. (2021) | Monthly | Sichuan Province in China | Artificial neural networks, geographically weighted regression, kriging, random forests |
| Nguyen et al. (2021) | Daily | South Korea | Random forests |
| Shen and Yong (2021) | Annual | China | Gradient boosting decision trees, random forests, support vector regression |
| Zhang et al. (2021) | Daily | China | Artificial neural networks, extreme learning machines, random forests, support vector regression |
| Chen et al. (2022a) | Daily | Coastal mountain region in the western United States | Deep learning |
| Fernandez-Palomino et al. (2022) | Daily | Ecuador and Peru | Random forests |
| Lin et al. (2022) | Daily | Three Gorges Reservoir area in China | Adaptive boosting decision trees, decision trees, random forests |
| Yang et al. (2022) | Daily | Kelantan river basin in Malaysia | Deep learning |
| Zandi et al. (2022) | Monthly | Alborz and Zagros mountain ranges in Iran | Artificial neural networks, locally weighted linear regression, random forests, stacked generalization, support vector regression |
| Militino et al. (2023) | Daily | Navarre in Spain | K-nearest neighbors, random forests, artificial neural networks |
| Present study | Monthly | Contiguous United States | Linear regression, multivariate adaptive regression splines, multivariate adaptive polynomial splines, random forests, gradient boosting machines, extreme gradient boosting, feed-forward neural networks, feed-forward neural networks with Bayesian regularization |

Machine learning for spatial interpolation has gained prominence in various fields of environmental science (Li et al. 2011). These fields include, but are not limited to, the agricultural sciences (Baratto et al. 2022), climate science (Sekulić et al. 2020b, Sekulić et al. 2021), hydrology (Tyralis et al. 2019c, Papacharalampous and Tyralis 2022b) and soil science (Wadoux et al. 2020, Chen et al. 2022b). Among the various machine learning algorithms, random forests seem to be the most frequently used ones (see the examples in Hengl et al. 2018). Notably, as machine learning algorithms do not model spatial dependence explicitly in their original form, efforts have been made to remedy this shortcoming, either directly (Saha et al. 2021) or indirectly (Behrens et al. 2018, Sekulić et al. 2020a, Georganos et al. 2021, Georganos and Kalogirou 2022). By exploiting spatial dependence information, the algorithms become more accurate.

As it has been noted earlier, machine learning algorithms constitute a major means for merging satellite products and ground-based measurements for obtaining precipitation data. However, their empirical properties are still not well known. This holds because most of the existing studies investigate a few algorithms, and because their investigations may be limited in terms of the length of the time periods examined and the size of the geographical areas examined. Large-scale benchmark tests and comparisons that involve, by definition, many algorithms and, at the same time, are conducted for long time periods and large geographical regions could be useful in providing directions on which algorithm to implement in specific settings of practical interest; thus, they have started to appear in other hydrological sub-disciplines. Relevant examples are available in Papacharalampous et al. (2019) and Tyralis et al. (2021).

In this study, we work towards filling the above-identified gap. More precisely, we compare a larger number of machine learning algorithms than usual (see Table 1) with respect to how accurate they are in providing estimates of total monthly precipitation in spatial interpolation settings by merging gridded satellite products and ground-based measurements. Also, the comparison is made for a long time period and for a large geographical area (again contrary to the most common strategy that appears currently in the literature), with this area also having a dense ground-based gauge network, thereby leading to trustable results for the monthly time scale. Moreover, proper evaluations are made according to theory and best practices from the field of statistics, with the methodological aspects developed in this endeavour contributing to the transfer of knowledge in the overall topic of spatial interpolation using machine and statistical learning algorithms.

The remainder of the paper is structured as follows: Section 2 describes the algorithms selected and the methodology followed for exploring the relevant regression setting. Section 3 presents the data and the validation procedure. Section 4 presents the results. Section 5 discusses the most important findings and provides recommendations for future research. Section 6 concludes the work.

## 2. Methods

### 2.1 Machine learning algorithms for spatial interpolation

Eight machine learning algorithms were implemented in this work for conducting spatial interpolation and were extensively compared with each other in the context of merging gridded satellite products and gauge-based measurements. In this section, we briefly describe these algorithms, while their detailed description can be found in Hastie et al. (2009), James et al. (2013) and Efron and Hastie (2016). Such a description is outside the scope of this work, as the implementations and documentations of the algorithms are already available in the R programming language. The R packages utilized are listed in Appendix A.

#### 2.1.1 Linear regression

A linear regression algorithm models the dependent variable as a linear weighted sum of the predictor variables (Hastie et al. 2009, pp 43–55). The algorithm is optimized with a squared error scoring function.

#### 2.1.2 Multivariate adaptive regression splines

The multivariate adaptive regression splines (MARS; Friedman 1991, 1993) model the dependent variable with a weighted sum of basis functions. The total number of basis functions (product degree) and associated parameters (knot locations) are automatically determined from the data. Herein, we implemented an additive model with hinge basis functions. The implementation was made with the default parameters.

#### 2.1.3 Multivariate adaptive polynomial splines

Multivariate adaptive polynomial splines (poly-MARS; Kooperberg et al. 1997, Stone et al. 1997) use piecewise linear splines to model the dependent variable in an adaptive regression procedure. Their main differences compared to MARS are that they require *"linear terms of a predictor to be in the model before nonlinear terms using the same* *predictor can be added"*, along with *"a univariate basis function to be in the model before a tensor-product basis function involving the univariate basis function can be in the model"* (Kooperberg 2022). In the present work, the poly-MARS model was implemented with the default parameters.

*2.1.4 Random forests*

Random forests (Breiman 2001a) are an ensemble of regression trees based on bagging (acronym for "bootstrap aggregation"). The benefits accompanying the application of this algorithm were summarized by Tyralis et al. (2019b), who also documented its recent popularity in hydrology with a systematic literature review. In random forests, a fixed number of predictor variables are randomly selected as candidates when determining the nodes of the regression tree. Herein, random forests were implemented with the default parameters. The number of trees was equal to 500.

*2.1.5 Gradient boosting machines*

Gradient boosting machines (gbm) are an ensemble learning algorithm. In brief, they iteratively train new base learners using the errors of previously trained base learners (Friedman 2001, Mayr et al. 2014, Natekin and Knoll 2013, Tyralis and Papacharalampous 2021). The final algorithm is essentially a sum of the trained base learners. Optimizations are performed by using a gradient descent algorithm and by adapting the loss function. The latter is the squared error scoring function in the implementation of this work. In the same implementation, the optimization's scoring function was the squared error and the base learners were regression trees. Also, the number of trees was set equal to 500 for keeping consistency with the implementation of the random forest algorithm. The defaults were used for the remaining parameters.

*2.1.6 Extreme gradient boosting*

Extreme gradient boosting (XGBoost; Chen and Guestrin 2016) is another boosting algorithm. It is considerably faster and better in performance in comparison to traditional implementations of boosting algorithms. It is also further regularized compared to such implementations for controlling overfitting. In the implementation of this work, the maximum number of the boosting iterations was set equal to 500. The remaining parameters were kept as default. For instance, the maximum depth of each tree was kept as equal to 6.

*2.1.7 Feed-forward neural networks*

Artificial neural networks (or simply "neural networks") extract linear combinations of the predictor variables as derived features and, subsequently, model the dependent variable as a nonlinear function of these features (Hastie et al. 2009, p 389). Herein, we used feed-forward neural networks (Ripley 1996, pp 143–180). The number of units in the hidden layer and the maximum number of iterations were set equal to 20 and 1 000, respectively, while the remaining parameters were kept as default.

*2.1.8 Feed-forward neural networks with Bayesian regularization*

Feed-forward neural networks with Bayesian regularization (MacKay 1992) for avoiding overfitting were also employed herein. In the respective implementation, the number of neurons that was set equal to 20 and the remaining parameters were kept as default. For instance, the maximum number of iterations was kept equal to 1 000.

## 2.2 Variable importance metric

We computed the random forests' permutation importance of the predictor variables, a metric measuring the mean increase of the prediction mean squared error on the out-of-bag portion of the data after permuting each predictor variable in the regression trees of the trained model and providing relative rankings of the importance of the predictor variables (Breiman 2001a). More generally, variable importance metrics can support explanations of the performance of machine learning algorithms (Breiman 2001b, Shmueli 2010), thereby expanding the overall scope of machine learning. This scope is often perceived as limited to the provision of accurate predictions. Random forests were fitted with 5 000 trees for computing variable importance.

## 3. Data and application

## 3.1 Data

Our experiments relied totally on open databases that offer earth-observed precipitation data at the monthly temporal resolution, gridded satellite precipitation data and elevation data for all the gauged locations and grid points shown in Figure 1.

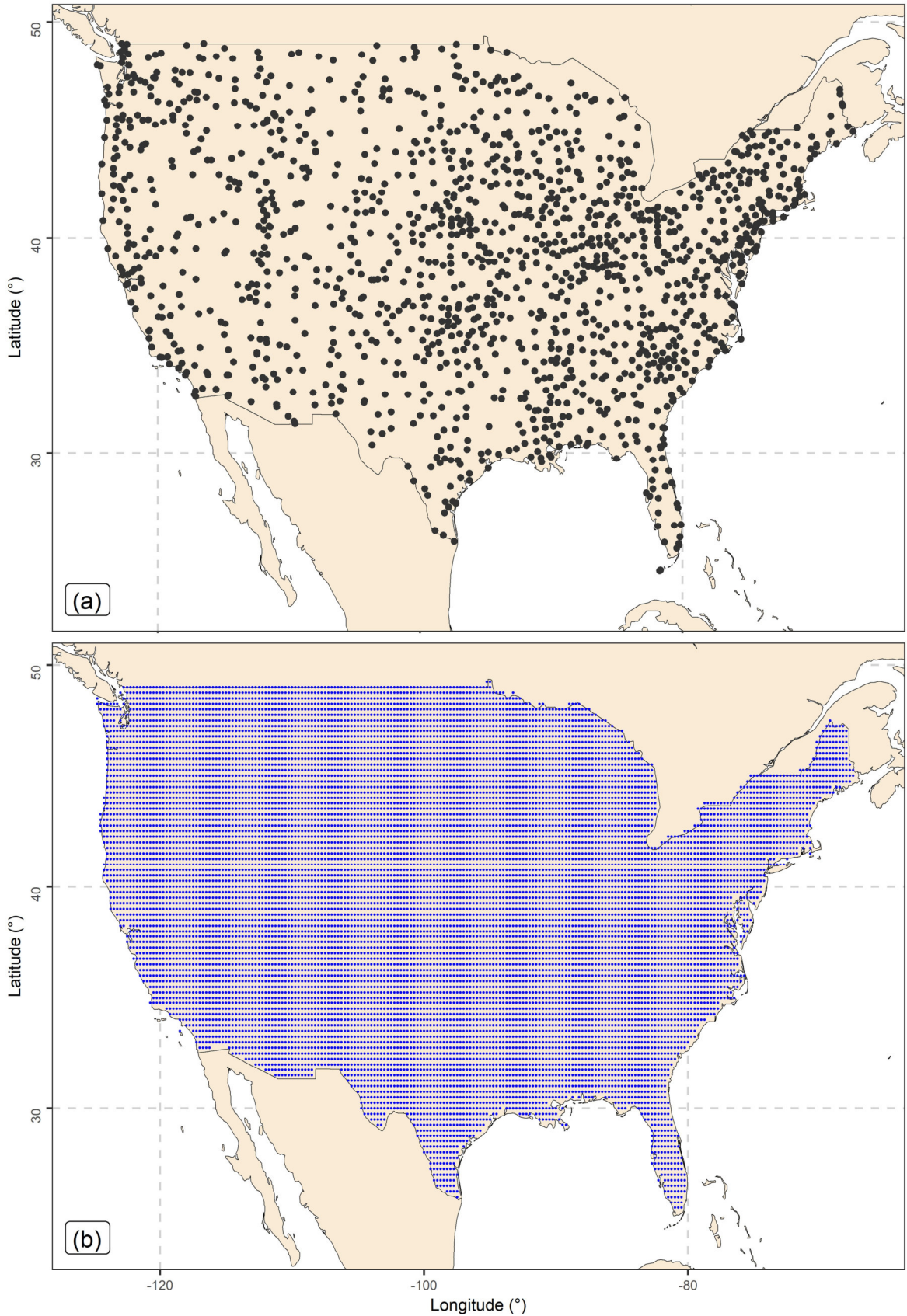

Figure 1. Maps of the geographical locations of: (a) the earth-located stations offering data for the present work; and (b) the points composing the PERSIANN grid defined herein.

### 3.1.1 Earth-observed precipitation data

Total monthly precipitation data of the Global Historical Climatology Network monthly database, version 2 (GHCNm; Peterson and Vose 1997) were used for the verification of the algorithms implemented for spatial interpolation. From the entire database, 1 421 stations that are located in the contiguous United States were extracted, and data that span the time period 2001–2015 were selected. These data were sourced from the website of the National Oceanic and Atmospheric Administration (NOAA) (https://www.ncei.noaa.gov/pub/data/ghcn/v2; accessed on 2022-09-24).

### 3.1.2 Satellite precipitation data

For the application, we additionally used precipitation data from the current operational PERSIANN (Precipitation Estimation from Remotely Sensed Information using Artificial Neural Networks) system. The latter was developed by the Centre for Hydrometeorology and Remote Sensing (CHRS) at the University of California, Irvine (UCI). The PERSIANN satellite data are created using artificial neural networks to establish a relationship between remotely sensed cloud-top temperature, measured by long-wave infrared (IR) sensors on geostationary orbiting satellites, and rainfall rates. Bias correction from passive microwave (PMW) records measured by low Earth-orbiting (LEO) satellites (Hsu et al. 1997, Nguyen et al. 2018, Nguyen et al. 2019) is also applied. These data were sourced in their daily format from the website of the Center for Hydrometeorology and Remote Sensing (CHRS) (https://chrsdata.eng.uci.edu; accessed on 2022-03-07).

The final product spans a grid with a spatial resolution of 0.25° x 0.25°. We extracted a grid that spans the contiguous United States at the time period 2001-2015. We also transformed daily precipitation into total monthly precipitation for supporting the investigations of this work.

### 3.1.3 Elevation data

For all the gauged geographical locations and the grid points shown in Figure 1, elevation data were computed by using the `get_elev_point` function of the `elevatr` R package (Hollister 2022). This function extracts point elevation data from the Amazon Web Services (AWS) Terrain Tiles (https://registry.opendata.aws/terrain-tiles; accessed on 2022-09-25). Elevation is a key variable in predicting hydrological processes (Xiong et al. 2022).

## 3.2 Validation setting and predictor variables

We define the earth-observed total monthly precipitation at the point of interest as the dependent variable. Notably, the ground-based stations are located irregularly in the region (see Figure 1); thus, the problem of defining predictor variables is not the usual one that is met in problems with tabular data. To form the regression settings, we found, separately for each station, the closest four grid points and we computed the distances $d_i$, $i$ = 1, 2, 3, 4 (in meters) from those points. We also indexed the points $S_i$, $i$ = 1, 2, 3, 4 according to their distance from the stations, where $d_1 < d_2 < d_3 < d_4$ (see Figure 2).

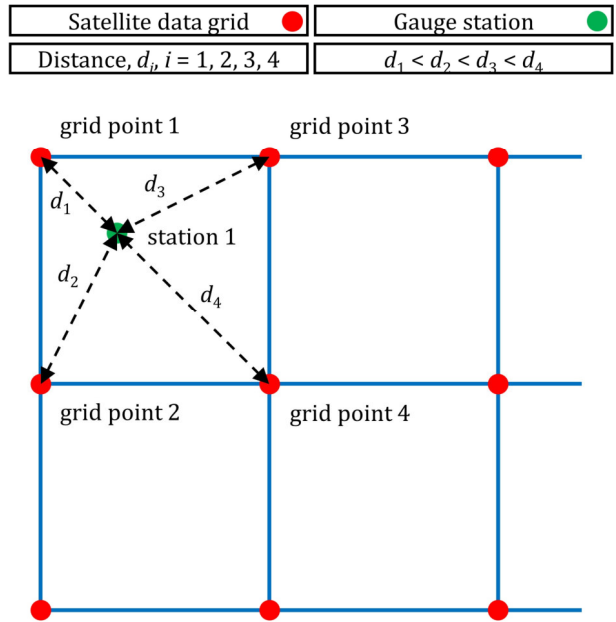

Figure 2. Setting of the regression problem. Note that the term "grid point" is used to describe the geographical locations with satellite data, while the term "station" is used to describe the geographical locations with ground-based measurements. Note also that, throughout the present work, the distances $d_i$, $i$ = 1, 2, 3, 4 are also referred to as "distances 1–4", respectively, and the total monthly precipitation values at the grid points 1–4 are referred to as "PERSIANN values 1–4", respectively.

Possible predictor variables for the technical problem of the present work are the total monthly precipitation values at the four closest grid points (which are referred to as "PERSIANN values 1–4"), the respective distances from the station (which are referred to as "distances 1–4"), the station's elevation, and the station's longitude and latitude. We defined and examined three different regression settings. Each of these correspond to a different set of predictor variables (see Table 2).

Table 2. Inclusion of predictor variables in the predictor sets examined in this work.

| Predictor variable | Predictor set 1 | Predictor set 2 | Predictor set 3 |
|---|---|---|---|
| PERSIANN value 1 | ✓ | ✓ | ✓ |
| PERSIANN value 2 | ✓ | ✓ | ✓ |
| PERSIANN value 3 | ✓ | ✓ | ✓ |
| PERSIANN value 4 | ✓ | ✓ | ✓ |
| Distance 1 | × | ✓ | ✓ |
| Distance 2 | × | ✓ | ✓ |
| Distance 3 | × | ✓ | ✓ |
| Distance 4 | × | ✓ | ✓ |
| Station elevation | ✓ | ✓ | ✓ |
| Station longitude | × | × | ✓ |
| Station latitude | × | × | ✓ |

The predictor sets 1 and 2 do not account directly for possible spatial dependences, as the station's longitude and latitude are not part of them. Still, by using these predictor sets, spatial dependence is modelled indirectly, through covariance information (satellite precipitation at close points and station elevation). The predictor set 2 includes more information with respect to the predictor set 1 and, more precisely, the distances between the station location and closest grid points. The predictor set 3 allows spatial dependence modelling, as it comprises the station's longitude and latitude.

The dataset is composed by 91 623 samples. Each sample includes the total monthly precipitation observation at a specific earth-located station for a specified month and a specified year, as well as the respective values of the predictor variables, with the latter being dependent on the regression setting (see Table 2). The results of the performance comparison are obtained within a five-fold cross-validation setting.

Overall, the validation setting proposed in this work benefits from the following:

− Stations with missing monthly precipitation values do not need to be excluded from the dataset, and missing values do not need to be filled. Instead, a varying number of stations are included in the procedure for each time point in the period investigated. In brief, we keep a dataset with the maximum possible size and we do not add uncertainties in the procedure by filling the missing values.

− The cross-validation is totally random with respect to both space and time. That is a standard procedure in the validation of precipitation products that combine satellite and earth-observed data.

− In the setting proposed, it is possible to create a corrected precipitation gridded dataset, because after fitting the regression algorithm it is possible to directly interpolate in the space conditional upon the predictor variables that are known.

– There is no need to first interpolate the station data to grid points and then verify the algorithms based on the earth-observed data previously interpolated. This procedure is common in the field, but it creates additional uncertainties.

A few limitations of the validation setting proposed in this work also exist. Indeed, there might be some degree of bias due to the fact that this setting does not incorporate, in a direct way, information on spatial dependencies. Such incorporations would require a different partitioning of the dataset (Meyer and Pebesma 2021, 2022), as machine learning models that may explicitly model spatial dependencies (see, e.g., Liu et al. 2022, Talebi et al. 2022) may not be applicable in settings with varying number of spatial observations at different times.

To deliver exploratory insight into the technical problem investigated in this work, we additionally estimated Spearman correlation (Spearman 1904) for all the possible pairs of the variables appearing in the regression settings. We also ranked the total of the predictor variables with respect to their importance in predicting the dependent variable. The latter was performed after estimating the importance according to Section 2.2.

## 3.3 Performance metrics and assessment

To compare the algorithms outlined in Section 2.1 in performing the spatial interpolation, we used the squared error scoring function. This function is defined by

$$S(x, y) := (x - y)^2 \qquad (1)$$

In the above equation, $y$ is the realization (observation) of the spatial process and $x$ is the prediction. The squared error scoring function is consistent for the mean functional of the predictive distributions (Gneiting 2011). Predictions of models in hydrology should be provided in probabilistic terms (see, e.g., the relevant review by Papacharalampous and Tyralis 2022a); still, a specific functional of the predictive distribution may be of interest. A model trained with the squared error scoring function predicts the mean functional of the predictive distribution (Gneiting 2011).

The performance criterion for the machine learning algorithms takes the form of the median squared error (MedSE) by computing the median of the squared error function, separately for each set {machine learning algorithm, predictor set, test fold}, according to Equation (2). In this equation, the subscript to $x$ and $y$, i.e., $i \in \{1, ..., n\}$, indicates the sample.

$$\text{MedSE} := \text{median}_n\{S(x_i, y_i)\} \tag{2}$$

The five MedSE values computed for each set {machine learning algorithm, predictor set} (with each corresponding to a different test fold) were then used to compute five relative scores (which are else referred to as "relative improvements" herein), separately for each predictor set, by using the set {linear regression, predictor set} as the reference case. These relative scores were then averaged, separately for each set {machine learning algorithm, predictor set}, to provide mean relative scores (which are else referred to as "mean relative improvements" herein). A skill score with linear regression as the reference technique for an arbitrary algorithm of interest $k$ is defined by

$$S_{\text{skill}} := \text{MedSE}_{\{k, \text{ predictor set}\}}/\text{MedSE}_{\{\text{linear regression, predictor set}\}} \tag{3}$$

The relative scores computed for the assessment are defined by

$$\text{RS}_{\{\text{linear regression, predictor set}\}} := 100\,(1 - S_{\text{skill}}) \tag{4}$$

To extend the comparison by also including the assessment between differences in performance across predictor sets, the procedures for computing the relative and mean relative scores were repeated by considering the set {linear regression, predictor set 1} as the reference case for all the sets {machine learning algorithm, predictor set}. In addition to the two types of relative improvements, we present information on the rankings of the machine learning algorithms. For obtaining the respective results, we first ranked the eight machine learning algorithms, separately for each set {case, predictor set} (with each case belonging to one test fold only). Then, we grouped these rankings per set {predictor set, test fold} and computed their mean. Lastly, we averaged the five mean ranking values corresponding to each predictor set and provided the results of this procedure, which are referred to in what follows as "mean rankings". Moreover, we repeated the mean ranking computation after computing the rankings collectively for all the predictor sets.

Notably, we did not compare the algorithms using alternative scoring functions (e.g., the absolute error scoring function) because such functions may not be consistent for the mean functional (excluding functions of the Bregman family; Gneiting [2011](#)). It is also possible to use other skill scores (e.g., the Nash-Sutcliffe efficiency, which is used widely in hydrology). Here, we preferred to use the simple linear regression algorithm as a reference technique. We believe that this choice is credible because of the simplicity and ease in the use of the algorithm.

## 4. Results

### 4.1 Regression setting exploration

Figure 3 presents the Spearman correlation estimates for all the possible pairs of the variables appearing in the settings examined in this work. The relationships between the predictand (i.e., the precipitation quantity observed at the earth-located stations) and the 11 predictor variables (see Section 3.2) can be assessed through the estimates displayed on the first column on the left side of the heatmap. Based on the Spearman correlation estimates, the strongest and, at the same time, equally strong among these relationships are those between the predictand and the four predictors referring to the precipitation quantities drawn from the PERSIANN grid. A possible explanation of this equality could be found in the Spearman correlation estimates made for the six pairs of PERSIANN values, which are equal to either 0.98 or 0.99, indicating extremely strong relationships. This strength can, in turn, be attributed to strong spatial relationships on the PERSIANN grid (i.e., to the fact that the neighbouring grid points have similar values) and, perhaps, also to the repetitions of values in the regression settings.

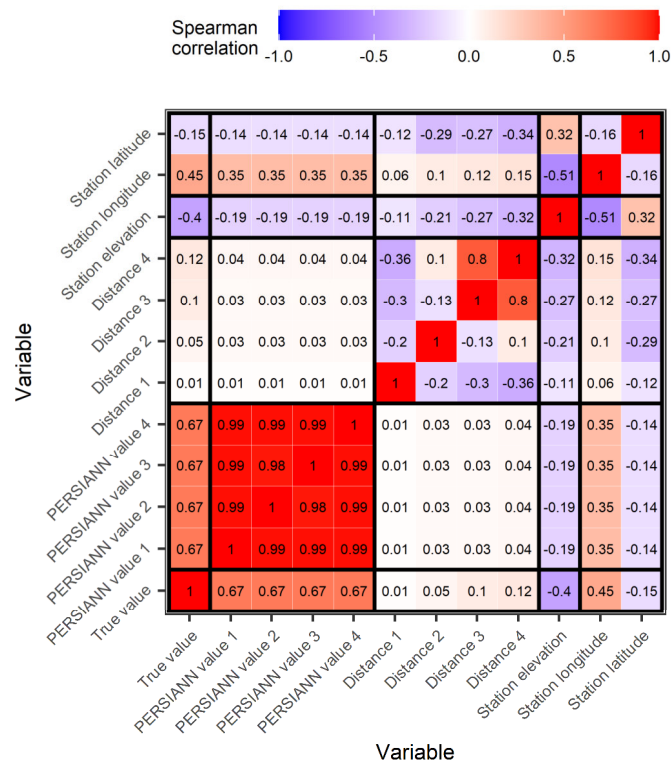

Figure 3. Heatmap of the Spearman correlation estimates for all the possible pairs of the variables appearing in the three regression settings.

Other relationships that are notably strong and, thus, expected, at least at an initial stage, to be particularly beneficial for estimating precipitation in the spatial setting adopted herein are those indicated by the values 0.45 and −0.40 (which again appear in the same column of the same heatmap; see [Figure 3](#)). The former of these two values refers to the relationship between the precipitation quantity observed at an earth-located station and the longitude at the location of this station, while the latter of the same values refers to the relationship between the precipitation quantity observed at an earth-located station and the elevation at the location of this station. The remaining relationships between the predictand and predictor variables are found to be less strong; nonetheless, they could also be worth-exploiting in the regression setting. Examples of the above-discussed relationships can be further inspected through [Figure 4](#).

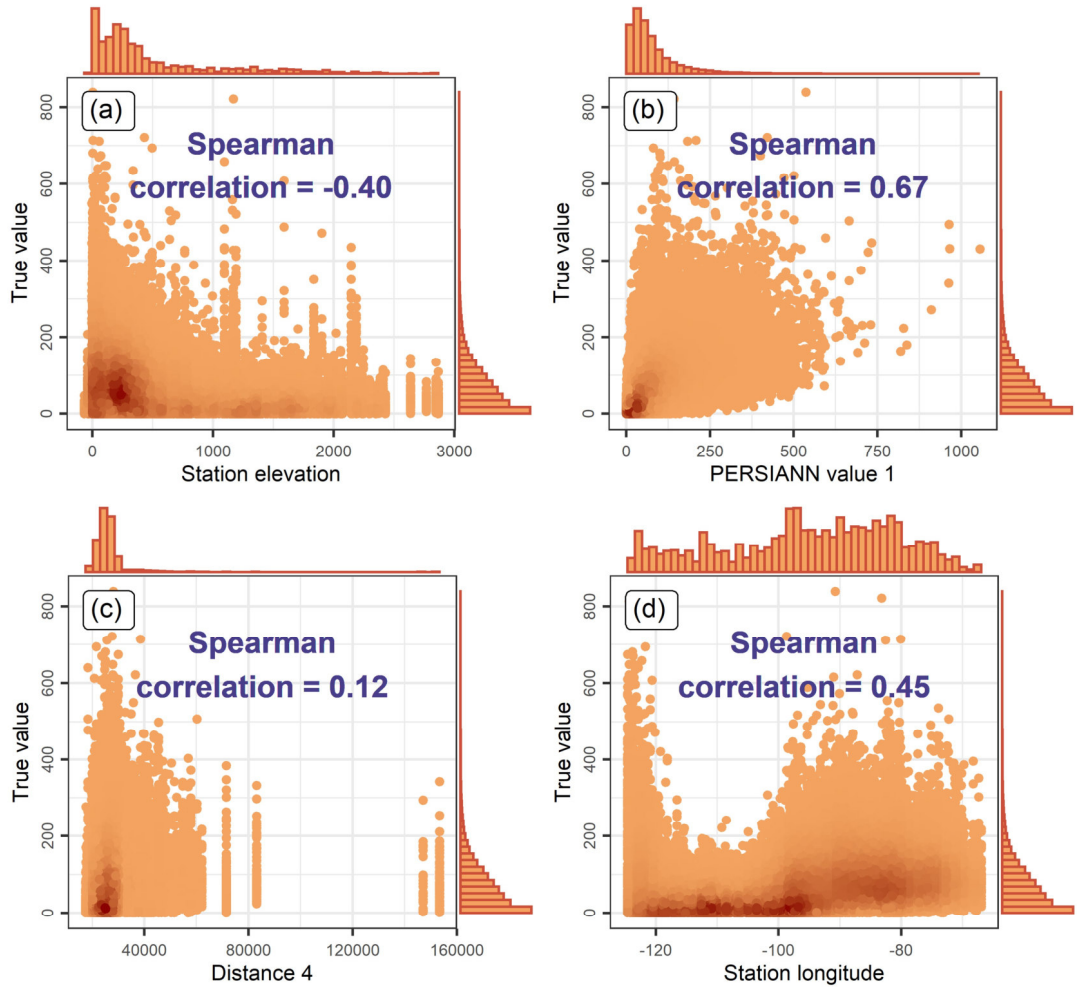

Figure 4. Scatterplots between the predictand (i.e., the precipitation value observed at an earth-located station) and the following predictor variables: (a) elevation at the location of this station; (b) precipitation value at the closest point on the PERSIANN grid for this station; (c) distance of the fourth closest point on the PERSIANN grid for this station; and (d) longitude at the location of this station. The Spearman correlation estimates are repeated here from Figure 3 for convenience. The redder the colour on the graphs, the denser the points.

Figure 5 presents the estimates of the importance of the 11 predictor variables; these estimates were provided by the random forest algorithm when considering all of these predictor variables in the regression setting. Figure 5 additionally provides the ordering of the same estimates, which is also the ordering of the 11 predictor variables according to their importance. The longitude at the location of the earth-located station is the most important predictor variable (probably because it is a spatial characteristic and the regression is made in a spatial setting), followed by the precipitation quantities drawn from the first, second and fourth closest points to the earth-located station at the PERSIANN grid. These latter three predictors are followed by the elevation at the location of the earth-located station. The next predictor in terms of importance is the precipitation quantity drawn from the third closest point to the earth-located station at the PERSIANN grid. The latitude at the location of the earth-located station follows and the four variables referring to distances are the least important ones.

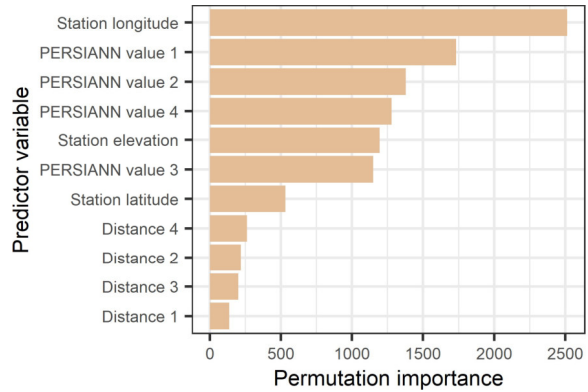

Figure 5. Barplot of the permutation importance scores of the predictor variables. The latter were ordered from the most to the least important ones (from top to bottom) based on the same scores.

## 4.2   Comparison of the algorithms

Figure 6 presents information that directly allows us to understand how the algorithms outlined in Section 2.1 performed with respect to each other in the various experiments, separately for each predictor set. Both the mean relative improvements (Figure 6a) and the mean rankings (Figure 6b) indicate that, overall, extreme gradient boosting (XGBoost) and random forests are the two best-performing algorithms. In terms of mean relative improvements, the former of these algorithms showed a much better performance than the latter when they were both run with the predictor sets 1 and 2, and a slightly better performance than the latter when they were both run with the predictor set 3. Feed-forward neural networks with Bayesian regularization follow in the line and, in terms of mean rankings, were empirically proven to have, an almost equally good performance with random forests. Multivariate adaptive polynomial splines (poly-MARS) and gradient boosting machines (gbm) are the fourth- and fifth-best-performing algorithms, respectively. While the mean rankings corresponding to the latter two algorithms do not suggest large differences in their performance, the mean relative improvements favour poly-MARS to a notable extent. In terms of both mean relative improvements and mean rankings, feed-forward neural networks performed better than gbm and multivariate adaptive regression splines (MARS) when these three algorithms were run with the predictor set 1. The linear regression algorithm was the worst for all the predictor sets investigated in this work. For the predictor sets 2 and 3, feed-forward neural networks were the second-worst algorithm with very close performance to that of linear regression, probably due to overfitting.

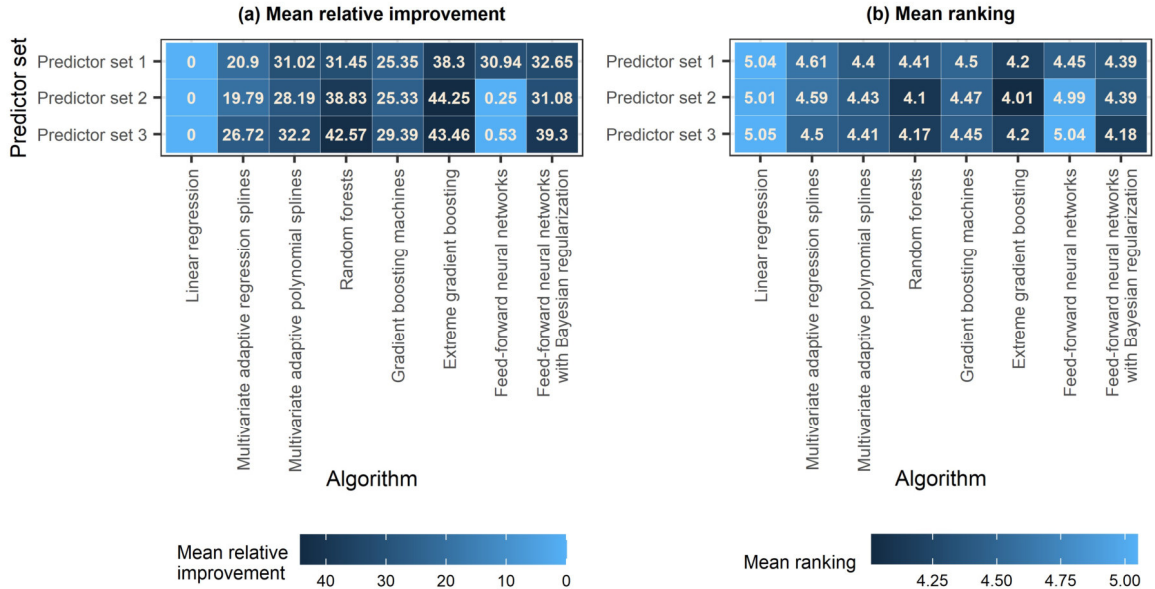

Figure 6. Heatmaps of: (a) the relative improvement (%) in terms of the median square error metric, averaged across the five folds, as this improvement was provided by each machine and statistical learning algorithm with respect to the linear regression algorithm; and (b) the mean ranking of each machine and statistical learning algorithm, averaged across the five folds. The computations were made separately for each predictor set. The darker the colour, the better the predictions on average.

Figure 7 facilitates comparisons, both across algorithms and across predictor sets, of the frequency with which each algorithm appeared in the various positions from the first to the eighth (i.e., the last) in the experiments. For the predictor set 1 (see Figure 7a), the linear regression algorithm was most commonly found in the last position, while its second most common position was the first and the six remaining positions appeared in much smaller and largely comparable frequencies. For the same predictor set, the XGBoost algorithm followed a notably similar pattern, although for it the first position was found to be the most common and the last position was found to be the second most common. The remaining positions appeared with smaller frequencies. In addition, the remaining algorithms were found less frequently in the first and last positions than the linear regression and XGBoost algorithms, with random forests appearing more often in these same positions than the other five algorithms. The frequency with which random forests appeared in the first, second, seventh and eighth positions is almost the same and larger than the frequency with which they appeared in the middle four positions. On the other hand, poly-MARS, feed-forward neural networks and feed-forward neural networks with Bayesian optimization appeared more often in the four middle positions than they appeared in the first two and last two positions, and MARS appeared more often in the six middle positions than it appeared in the first and last positions.

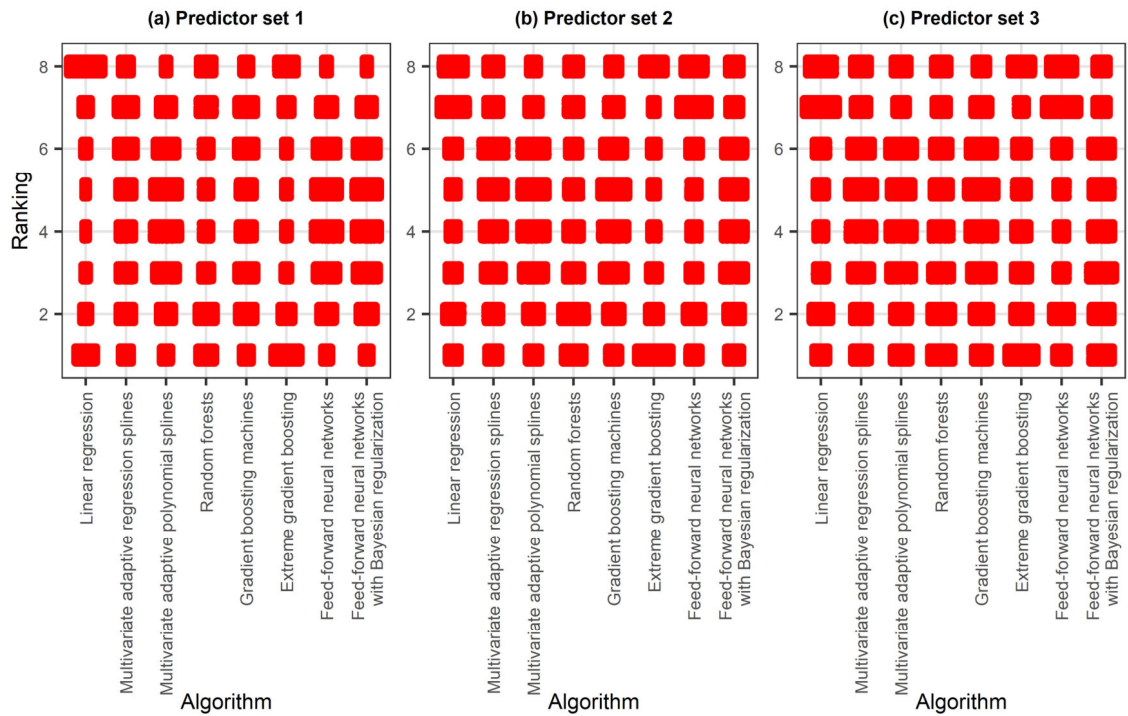

Figure 7. Sinaplots of the rankings from 1 to 8 of the machine and statistical learning algorithms for the predictor sets (a–c) 1–3. These rankings were computed separately for each pair {case, predictor set}.

For the predictor set 2 (see Figure 7b), there is differentiation in most of the above-discussed patterns. Notably, for this predictor set, the patterns observed for feed-forward neural networks and linear regression are quite similar. These algorithms appeared in one of the last two positions more often than any other algorithm. Moreover, the seventh position was more frequent for them, and their frequency of appearance in the first, third, fourth, fifth and sixth positions was almost the same and a bit smaller than their frequency of appearance in the second position. The same algorithms appeared in the last position equally often with the XGBoost algorithm. The latter is the algorithm that appeared most often in the first position by far. Similarly to what was previously noted for the predictor set 1, this algorithm appeared more frequently in the first and last position than in any other position for the predictor set 2, with the first position also being much more frequent than the last one. Random forests appeared more often in the first two positions than in any other position and the remaining algorithms appeared more often in the third, fourth, fifth and sixth positions than in the remaining four positions.

For the predictor set 3 (see Figure 7c), the frequency with which each algorithm appeared in the various positions from the first to the last exhibits more similarities with what was found for the predictor set 2 than with what was found for the predictor set 1. Yet, there are a few notable differences with respect to this good reference case. In fact, although the XGBoost algorithm appeared more often, here as well, in the first and last positions, the frequency of its appearance in the remaining positions was notably larger than the respective frequency for the case of the predictor set 2. In addition, the random forest algorithm appeared more often in the third, fourth, fifth and sixth positions than it did for the same reference case.

Moreover, Figures 8 and 9 allow us to understand how much the additional predictors in the predictor sets 2 and 3 improved or deteriorated the performance of the eight algorithms with respect to using the predictor set 1. The computed improvements were found to be all positive and particularly large for the random forest and the two boosting algorithms, especially when moving to the predictor set 3. Also notably large and positive are the performance improvements offered by the additional predictors in the predictor set 3 with respect to the predictor set 1 for linear regression, MARS, poly-MARS and feed-forward neural networks with Bayesian regularization, while the same does not apply for the case of using the predictor set 2 instead of the predictor set 1 for the same algorithms. Figure 8 further reveals the best-performing combinations of algorithms and predictors. These are the {extreme gradient boosting, predictor set 3} and {random forests, predictor set 3}, with the former offering slightly better performance in terms of mean relative improvement (but not in terms of mean ranking).

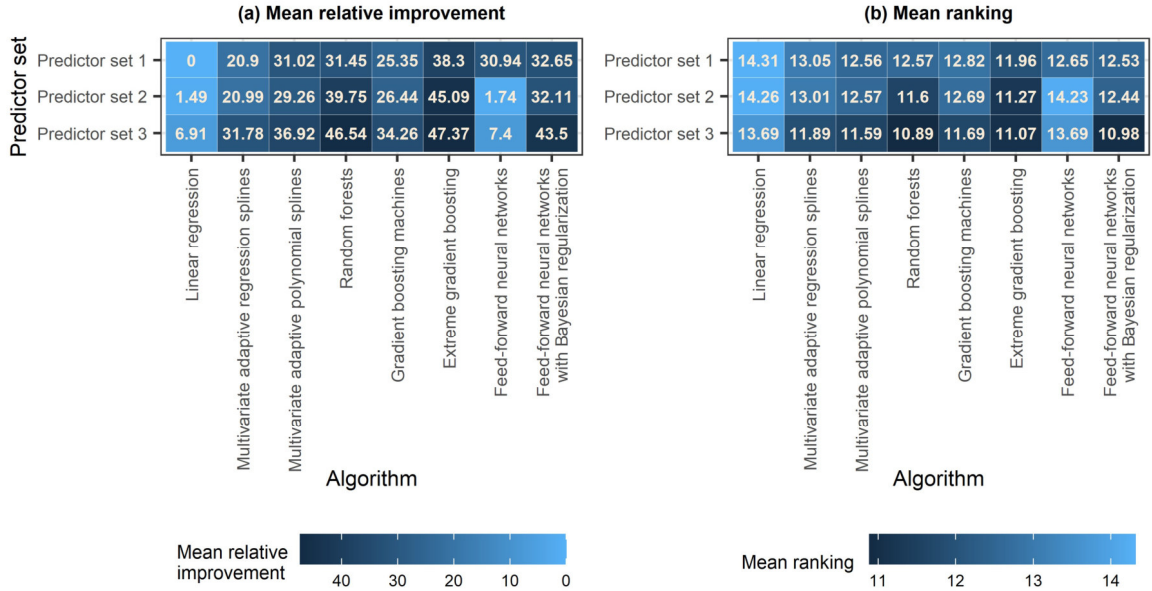

Figure 8. Heatmaps of: (a) the relative improvement (%) in terms of the median square error metric, averaged across the five folds, as this improvement was provided by each machine and statistical learning algorithm with respect to the linear regression algorithm, with this latter algorithm being run with the predictor set 1; and (b) the mean ranking of each machine and statistical learning algorithm, averaged across the five folds. The computations were made collectively for all the predictor sets. The darker the colour, the better the predictions on average.

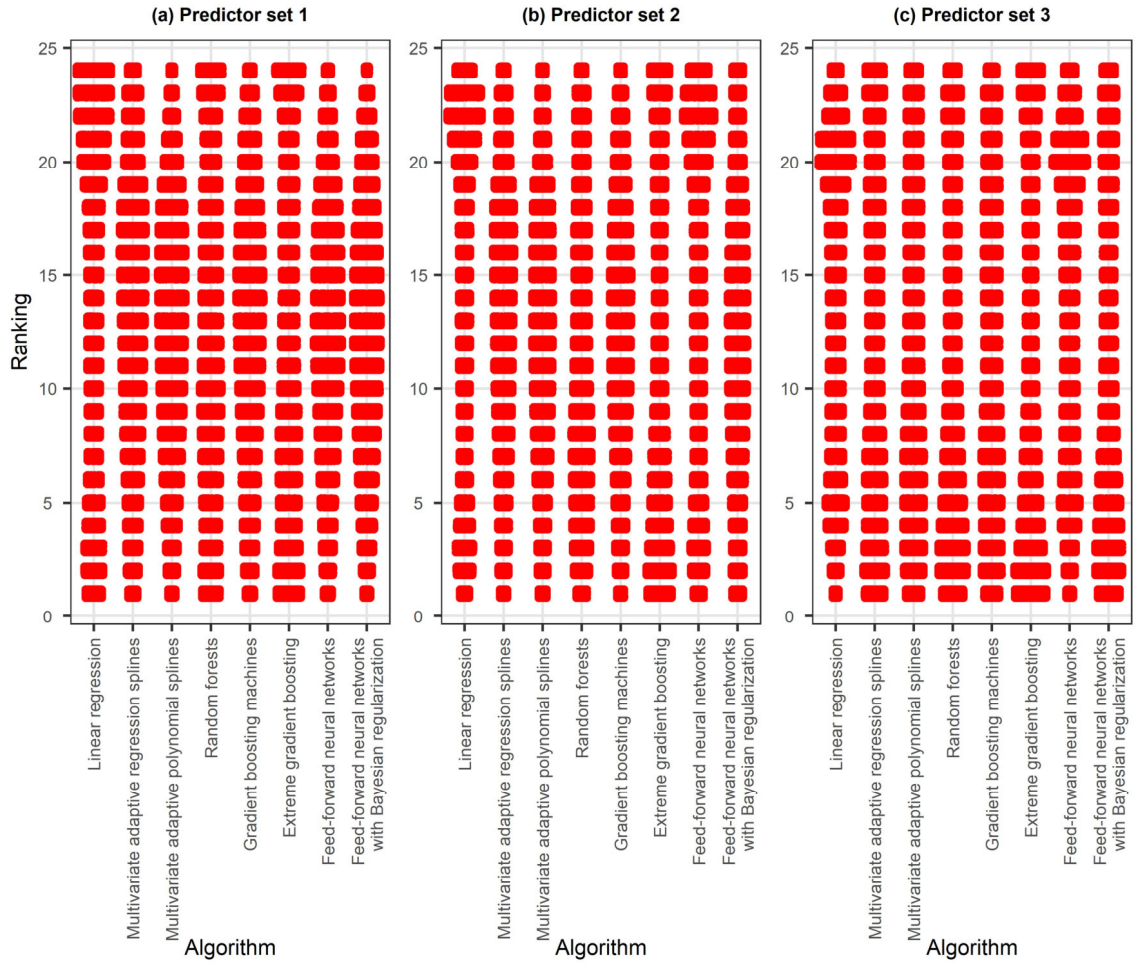

Figure 9. Sinaplots of the rankings from 1 to 24 of the machine and statistical learning algorithms for the predictor sets (a–c) 1–3. These rankings were computed separately for each case and collectively for all the predictor sets.

Perhaps it is also relevant to highlight, at this point, that the combination {feed-forward neural networks with Bayesian regularization, predictor set 3} was in the fourth position in terms of mean relative improvement (surpassing all the remaining combinations aside from the two best-performing ones and the {extreme gradient boosting, predictor set 2}; see Figure 8a) and in the second position in terms of mean ranking (surpassing all the remaining combinations aside from the {random forests, predictor set 3}; see Figure 8b). At the same time, according to Figure 8a, the feed-forward neural networks without Bayesian regularization performed so poorly when applied with the predictor sets 2 and 3 (in which the number of the predictor variables increases by four and six, respectively, with respect to the predictor set 1), that they were only slightly better than the linear regression model when applied with the predictor sets 2 and 3, respectively. Lastly, according to the same figure, the combination {linear model, predictor set 3} outperformed the combination {feed-forward neural networks, predictor set 2}.

## 5. Discussion

In summary, the large-scale comparison showed that the machine learning algorithms of this work can be ordered from the best to the worst in regard to their accuracy in correcting satellite precipitation products at the monthly temporal scale as follows: extreme gradient boosting (XGBoost), random forests, Bayesian regularized feed-forward neural networks, multivariate adaptive polynomial splines (poly-MARS), gradient boosting machines (gbm), multivariate adaptive regression splines (MARS), feed-forward neural networks, linear regression. The differences in performance were found to be smaller between some pairs of algorithms when the application is made with specific predictors (e.g., random forests and XGBoost when run with the predictor set 3) and larger (or medium) in other cases. Especially the magnitude of the differences computed between each of the two best-performing and the remaining algorithms, for the case in which the most information-rich predictor set is exploited, suggests that the consideration of the findings of this work can have a large positive impact on future applications. Notably, the fact that the random forest, XGBoost and gbm algorithms perform better or, in the worst case, similarly when predictors are added could be attributed to their known theoretical properties. Summaries of these properties are provided in the reviews by Tyralis et al. (2019b) and Tyralis and Papacharalampous (2021), where extensive lists of references to the related machine learning literature are also provided.

Aside from the selection of a machine learning algorithm and the selection of a set of predictor variables, which are well-covered by this work for the monthly temporal scale, there are also other important themes, whose investigation could substantially improve performance in the problem of correcting satellite precipitation products at the various temporal scales. Perhaps the most worthy of discussion here is the use of ensembles of machine learning algorithms in the context of ensemble learning. A few works are devoted to ensemble learning algorithms for spatial interpolation (e.g., Davies and Van Der Laan 2016, Egaña et al. 2021) and could provide a starting point, together with the present work, for building detailed big data comparisons of ensemble learning algorithms. Note here that the ensemble learning algorithms include the simple combinations (see, e.g., those in Petropoulos and Svetunkov 2020, Papacharalampous and Tyralis 2020) and more advanced stacking and meta-learning approaches (see, e.g., those in Wolpert 1992;

Tyralis et al. 2019a, Montero-Manso et al. 2020, Talagala et al. 2021), and are increasingly adopted in many fields, including hydrology.

Other possible themes for future research, in the important direction of improving both our understanding of the practical problem of correcting satellite precipitation products and the various algorithmic solutions to this problem, include the investigation of spatial and temporal patterns (as the precipitation product correction errors might follow such patterns) and the explanation of the predictive performance of the various algorithms by combining time series feature estimation (see multiple examples of time series features in Fulcher et al. 2013, Kang et al. 2017) and explainable machine learning (see, e.g., the relevant reviews in Belle and Papantonis 2021, Linardatos et al. 2021). Examples of such investigations are available for a different modelling context in Papacharalampous et al. (2022). Lastly, the comparisons could be extended to include algorithms for predictive uncertainty quantification. A few works are devoted to such machine learning algorithms for spatial interpolation (e.g., Fouedjio and Klump 2019). Still, comparison frameworks and large-scale results for multiple algorithms are currently missing from the literature of satellite precipitation data correction.

## 6. Conclusions

Hydrological applications often rely on gridded precipitation datasets from satellites, as these datasets cover large regions with higher spatial density compared to the ones that comprise ground-based measurements. Still, the former datasets are less accurate than the latter, with the various machine learning algorithms consisting an established means for improving their accuracy in regression settings. In these settings, the ground-based measurements play the role of the dependent variable and the satellite data play the role of the predictor variables, together with data for topography factors (e.g., elevation). The studies devoted to this important endeavour are numerous; still, most of them involve a limited number of machine learning algorithms, and are also conducted for a small region and a limited time period. Thus, their results are mostly of local importance, and cannot support the derivation of more general guidance and best practices.

In this work, we moved beyond the above-outlined standard approach by comparing eight machine learning algorithms in correcting precipitation satellite data for the entire contiguous United States and over a 15-year period. More precisely, we exploited monthly precipitation satellite data from the PERSIANN (Precipitation Estimation from Remotely

Sensed Information using Artificial Neural Networks) gridded dataset and monthly earth-observed precipitation data from the Global Historical Climatology Network monthly database, version 2 (GHCNm), and based the comparison on the squared error scoring function. Overall, extreme gradient boosting (XGBoost) and random forests were found to be the most accurate algorithms, with the former being more accurate than the latter to a small extent for the majority of the scores computed. The remaining algorithms can be ordered from the best- to the worst-performing as follows: feed-forward neural networks with Bayesian regularization, multivariate adaptive polynomial splines (poly-MARS), gradient boosting machines (gbm), multivariate adaptive regression splines (MARS), feed-forward neural networks, linear regression.

Aside from the above ordering which constitutes, in our opinion, the most important finding of the present work, important findings on the selection of predictor variables in the field of satellite precipitation data correction were also obtained for the monthly time scale. Indeed, we found that the distances of the four closest grid points from a ground-based station, as well as this station's longitude and latitude, can offer improvements in predictive performance, when utilized as predictor variables for most of the machine learning algorithms assessed (including the best-performing ones), together with the monthly precipitation values at the four closest grid points and the station's elevation. Also importantly, we proposed a new validation setting that could bring considerable benefits to future comparisons of machine and statistical learning algorithms in the field. These benefits are enumerated in Section 3.2. Even more generally, we proposed an authentic methodological framework and contributed to the transfer of theory and best practices from the field of statistics to the field of satellite precipitation data correction.

**Conflicts of interest:** The authors declare no conflict of interest.

**Author contributions:** GP and HT conceptualized and designed the work with input from AD and ND. GP and HT performed the analyses and visualizations, and wrote the first draft, which was commented on and enriched with new text, interpretations and discussions by AD and ND.

**Funding:** This work was conducted in the context of the research project BETTER RAIN (BEnefiTTing from machine lEarning algoRithms and concepts for correcting satellite RAINfall products). This research project was supported by the Hellenic Foundation for Research and Innovation (H.F.R.I.) under the "3rd Call for H.F.R.I. Research Projects to support Post-Doctoral Researchers" (Project Number: 7368).

**Acknowledgements:** The authors are sincerely grateful to the Journal for inviting the submission of this paper, and to the Editor and Reviewers for their constructive remarks. They would also like to acknowledge the contribution of the late Professor Yorgos Photis in the proposal of the research project BETTER RAIN.

## Appendix A  Statistical software information

We used the `R` programming language (R Core Team 2022) to implement the algorithms, and to report and visualize the results.

For data processing and visualizations, we used the contributed `R` packages `caret` (Kuhn 2022), `data.table` (Dowle and Srinivasan 2022), `elevatr` (Hollister 2022), `ggforce` (Pedersen 2022), `ncdf4` (Pierce 2021), `rgdal` (Bivand et al. 2022), `sf` (Pebesma 2018, 2022), `spdep` (Bivand 2022, Bivand and Wong 2018, Bivand et al. 2013), `tidyverse` (Wickham et al. 2019, Wickham 2022).

The algorithms were implemented by using the contributed `R` packages `brnn` (Rodriguez and Gianola 2022), `earth` (Milborrow 2021), `gbm` (Greenwell et al. 2022), `nnet` (Ripley 2022, Venables and Ripley 2002), `polspline` (Kooperberg 2022), `ranger` (Wright 2022, Wright and Ziegler 2017), `xgboost` (Chen et al. 2022c).

The performance metrics were computed by implementing the contributed `R` package `scoringfunctions` (Tyralis and Papacharalampous 2022a, 2022b).

Reports were produced by using the contributed `R` packages `devtools` (Wickham et al. 2022), `knitr` (Xie 2014, 2015, 2022), `rmarkdown` (Allaire et al. 2022, Xie et al. 2018, 2020).